\documentclass[prl,superscriptaddress,twocolumn]{revtex4-1}
\usepackage[final]{graphicx}
\usepackage{times,bbm,amsmath,amssymb}
\usepackage{epsfig,color}
\usepackage{hyperref}
\usepackage{float,siunitx}
\usepackage[caption = false]{subfig}
\usepackage[greek,english]{babel}
\usepackage{thumbpdf,enumerate}
\usepackage{booktabs}
\usepackage{sidecap}
\usepackage[scaled=.8]{couriers}
\usepackage{pstricks}
\usepackage{multirow}
\usepackage{placeins}
\usepackage{pst-grad,bm}
\usepackage{epigraph}
\usepackage{gensymb}
\usepackage{longtable}
\usepackage{booktabs}
\usepackage{gensymb}

\def\Tr{\hbox{Tr}} 
\usepackage{pifont}

\newcommand{\ket}[1]{\vert#1\rangle}
\newcommand{\bra}[1]{\langle#1\vert}

\newcommand{\sys}{\cal S}
\newcommand{\env}{\cal E}
\newcommand{\frag}{\cal F}

\begin{document}

\title{Experimental signature of Quantum Darwinism in photonic cluster states}
\author{Mario A. Ciampini}\email{marioarnolfo.ciampini@uniroma1.it}
\affiliation{Dipartimento di Fisica, Sapienza Universit\`a di Roma, P.le Aldo Moro 5, 00185, Rome, Italy}
\affiliation{University of Vienna, Faculty of Physics, Vienna Center for Quantum
Science and Technology (VCQ), Boltzmanngasse 5, 1090 Vienna, Austria}

\author{Giorgia Pinna}
\affiliation{Dipartimento di Fisica, Sapienza Universit\`a di Roma, P.le Aldo Moro 5, 00185, Rome, Italy}

\author{Paolo Mataloni}
\affiliation{Dipartimento di Fisica, Sapienza Universit\`a di Roma, P.le Aldo Moro 5, 00185, Rome, Italy}

\author{Mauro Paternostro} 
\affiliation{Centre for Theoretical Atomic, Molecular and Optical Physics, School of Mathematics and Physics, Queen's University Belfast, Belfast BT7 1NN, United Kingdom}

\begin{abstract}
We report on an experimental assessment of the emergence of Quantum Darwinism (QD) from engineered open-system dynamics. We use a photonic hyperentangled source of graph states to address the effects that correlations among the elements of a multi-party environment have on the establishment of objective reality ensuing the quantum-to-classical transition. Besides embodying one of the first experimental efforts towards the chaarcterization of QD, our work illustrates the non-trivial consequences that multipartite entanglement and, in turn, the possibility of having environment-to-system back-action have on the features of the QD framework.
\end{abstract}

\maketitle

\emph{Introduction. --} The field of quantum open systems attempts at shedding light on the process translating the non-classical state of a system all the way down to a mundane classical description~\cite{Schlosshauer,Zurek,Breuer}. The origin of such a fundamental mechanism - which is dubbed {\it quantum-to-classical transition} - and the features of its occurrence remain not completely clear or understood, and are the topic of much research effort~\cite{Schlosshauer2, Bassi}. One of the most accredited explanation relies on the action of environment-induced decoherence~\cite{Zurek2}, a phenomenon where the environment surrounding a given system of interest continuously monitors the state of the latter, thus acquiring information about it. The consequence of the environmental monitoring process is that {\it fragile} quantum superpositions of the system under scrutiny are removed in time, while classical mixtures of {\it more robust} macroscopic states survive, leading to the transition to classsicality~\cite{Zurek3}. 

Very often, in light of the typical assumption of an environment consisting of a very large number of subsystems whose dynamics is virtually impossible to track, the centres of attention are the system itself and its properties. The environment is thus eliminated from the dynamical picture, and the effect of its coupling to the system retained "effectively" in the properties of the ensuing system's non-unitary evolution. However, much can be learned from the retention of the state of the environment and the shift of the attention towards the features of the information-acquisition process that is at the basis of environment-induced decoherence. 

This is precisely what the framework of Quantum Darwinism (QD) aims at doing by putting the environment back into the description of the dynamics~\cite{Zurek4}. Consider a quantum system $S$ interacting with an environment $\mathcal{E}$ made up of many independent (and mutually non-interacting) subsystems $\mathcal{E}^{(j)}$ through an information-transfer process.
External observers that want to get information about ${\cal S}$ are allowed to do so only by directly measuring ${\cal E}^{(j)}$. In the QD picture the objective, classical description of ${\cal S}$ emerges from the quantum one due to the proliferation of \textit{redundant} information throughout the environment resulting from the ${\cal S}$-${\cal E}$ correlations established by their mutual interaction. More precisely, key to QD is that the states produced by environment-induced decoherence encode many local copies of classical information about ${\cal S}$. Such a proliferation allows information about the system to be extracted from different fragments of the environment, and intercepted by the observers. The larger the number of fragments that acquire information $\sim H(\rho_{\cal S})$ about the state $\rho_{\cal S}$ of the system (here $H(\rho_{\cal S})=-{\rm Tr}[\rho\log_2\rho]$ is the von Neumann entropy), the more widespread is the classical data recorded by the elements of the environment.
The phenomenology of QD is yet to be fully characterized and it is in general very interesting to determine fully the domain of validity of its framework~\cite{Brandao}.

The goal of this work is to experimentally address the possible effects of intra-environment correlations in the occurrence of QD, thus addressing a situation that deviates significantly form the typical assumption of independent ${\cal E}_j$'s made within the QD framework.
 We build our analysis on the use of specifically tailored multi-qubit graph states~\cite{Barbieri,Cinelli,Barreiro,Vallone,Chen}, within which we identify a system qubits and environmental ones. Graph states result from the evolution of an $N$-partite register of qubits initially prepared in state $\otimes^N_{j=1}\hat H_j\ket{+}_j$, with $\ket{+}$ the eigenstate of the Pauli $\sigma_x$ operator with eigenvalue $+1$H, following the Hamiltonian of interaction  \begin{equation}
\hat H=\frac14\sum^N_{j,k=1}{g_{jk}}(\openone-\sigma^j_{z})(\openone-\sigma^k_z),
 \end{equation}
 where $g_{jk}$ is a coupling rate (we assume units such that $\hbar=1$ throughout this work). The degree of correlations shared by any two qubits in the register is a function of the rate $g_{jk}$. In our investigation we tune such parameters from $g_{jk}=0$, when the indices $j$ and $k$ pertain to environmental elements only, to the case of non-null intra-environment interactions. Therefore, we compare explicitly the case of independent sub-environments (in line with the QD assumptions) to that where strong intra-environment correlations are set.  The corresponding analysis of the occurrence of QD shows that significant deviations from the expectations for such phenomenon are in order in the latter case, with seemingly no redundant information about the state of the chosen ${\cal S}$ being recorded by the environment.
 

\begin{figure}[t!]
\includegraphics[width=\columnwidth]{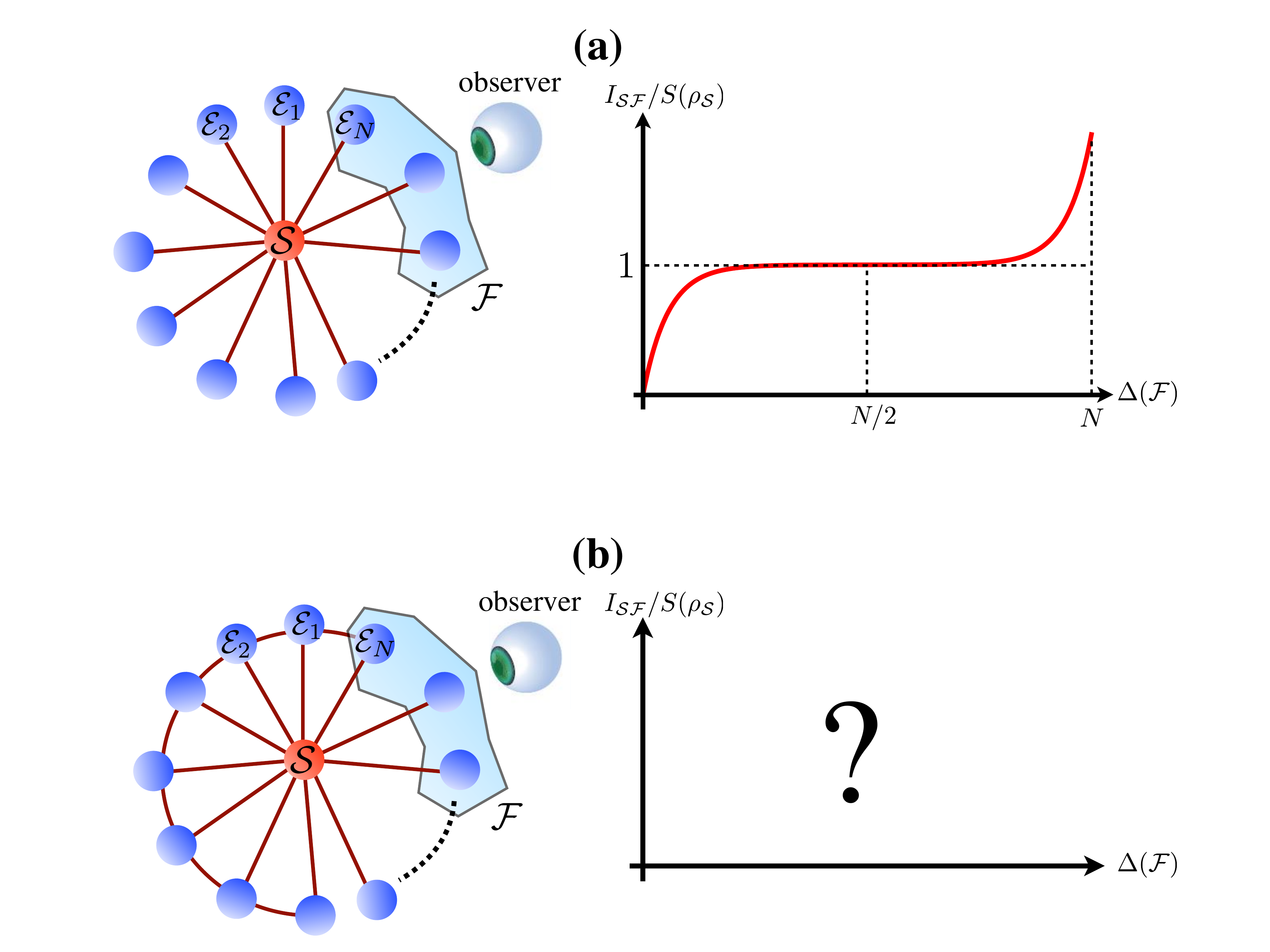}
\caption{Upper panel: under the assumptions of QD (independent sub-environments ${\cal E}_j$), the mutual information $I_{\sys\frag}$ between the system $\sys$ and a fragment $\frag$ of the environment showcases a {\it plateau} against the size $\Delta(\frag)$ of the fragment itself, which witnesses the redundancy of the information encoded in the state of the elements of the environment. Lower panel: under the presence of intra-environment correlations (indicated by the solid lines joining pairs of ${\cal E}_j$ particles), the phenomenology of QD is not known.}
\label{fig:logic}
\end{figure}

\emph{Brief description of QD. --} Let us assume, for the sake of simplicity, that the initial state of the system reads $\ket{\psi}_{\sys}= \sum_{k=1}^{n}\psi_k \ket{s_k}_{\cal{S}}$ with $\{\ket{s_k}_{\sys}\}$ a basis in the Hilbert space of the $n$-dimensional system. The environment is prepared in $\ket{\epsilon_0}_{\env}=\otimes^N_{j=1}\ket{\epsilon_0}_j$ with $\ket{\epsilon_0}_j$ the initial state of the $j^\text{th}$ sub-environment. The paradigm of QD requires the following typical evolution of the initial $\sys-\env$ state 
\begin{equation}
\ket{\psi}_{\sys}\ket{\epsilon_0}_{\env}\longrightarrow
\sum_{k=1}^{n}\psi_k \ket{s_k}_{\sys}\ket{\epsilon_k}_{\env}
\end{equation}
with 
$\ket{\epsilon_k}_{\env}=\otimes^N_{j=1}\ket{\epsilon_k}_j$ the evolved state of the environment conditional on the system being in $\ket{s_k}_{\sys}$.  
As all the sub-environments encode the same state and there is strong correlation between $\sys$ and $\env$, we can say that the information on the system is reduntantly recorded into the environment (such information being the {\it classical} label $k$ that identifies the state component of the system). 

The redundancy that is at the core of the QD phenomenology is well captured by the degree of total correlations set in the joint state of $\sys$ and a fragment ${\cal F}$ of the environment, i.e. the set of sub-environments corresponding to a given choice of the indices $j=1,\dots,N$. Such total correlations are quantified by the mutual information 
\begin{equation}
I_{\sys\frag}= H_{\sys} +H_{\frag}- H_{\sys,\frag}.
\end{equation}
The emergence of QD, associated with the redundant proliferation of information across $\env$, is marked by the insensitivity of $I_{\sys\frag}$ of the dimension $\Delta(\frag)$ of the fragment $\frag$ being considered.  
This happens when almost all of the information about $\sys$ is contained in  $\frag$, so that $I_{\sys\frag}$ quickly rises to $S_{\sys}$, which is all of the available information about $\sys$ [cf. Fig.~\ref{fig:logic}]. The mutual information $I_{\sys\frag}$ is the main instrument of the analysis that we present hereafter. 

\begin{figure}[t!]
{\bf (a)}\hskip3.5cm{\bf (b)}
\includegraphics[width=0.9\columnwidth]{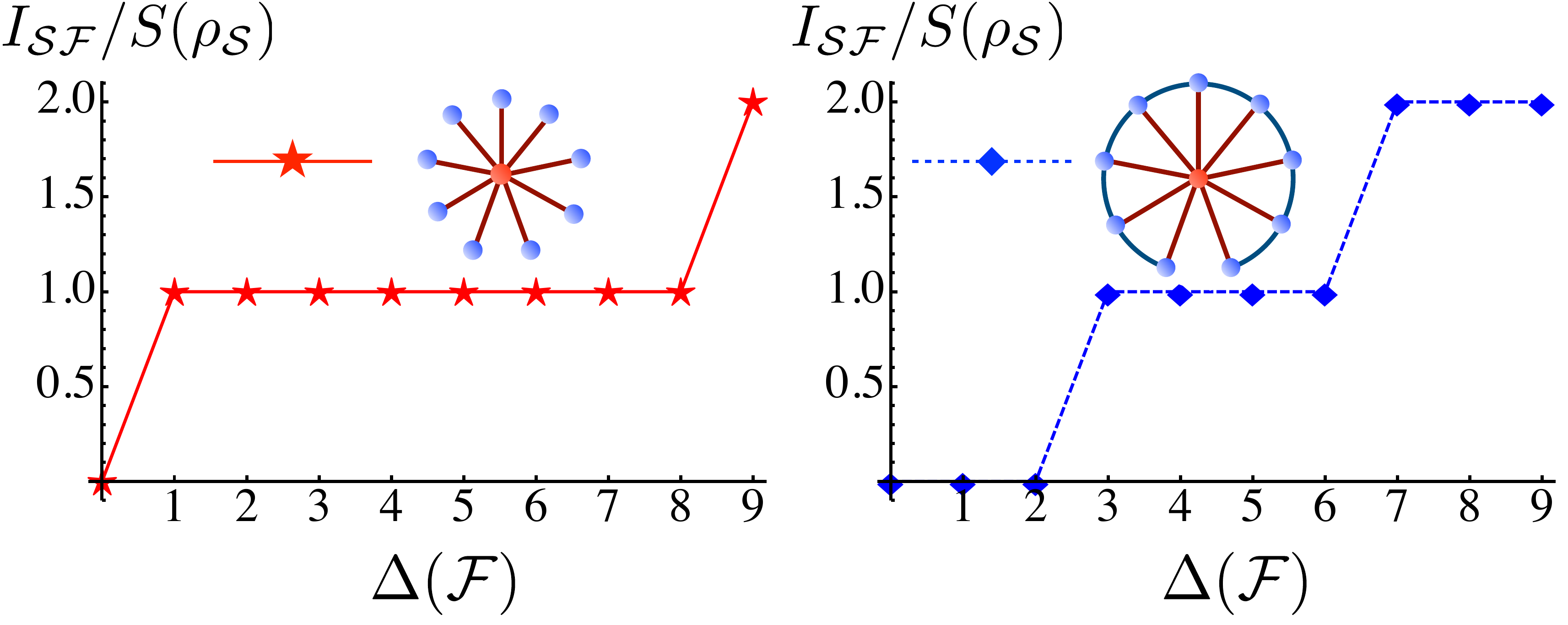}
\caption{
Numerical analysis of the mutual information between the system and fragments of the environment as a function of the number of qubits in the fragment. The results are shown for a ten-element star-shaped [panel {\bf (a)}] and diamond-shaped [panel {\bf (b)}] graph state. We include the underlying graph of the states that have been considered in the calculations, which involve a system qubit $\sys$ and nine elements of its environment $\env$. The burgundi-colored link are the interactions set between $\sys$ and ${\cal E}_j$, while the blue-colored ones stand for the intra-environment interactions.}
\label{fig:graph01}
\end{figure}

\emph{Resource state for the study of QD and its break-down. --} As anticipated above, our analysis will be based on the use of graph states able to encode tuneable correlations between the system and the environmental elements ${\cal E}_j$. Specifically, we consider states whose underlying graphs are akin to the those shown in Fig.~\ref{fig:graph01}, and which we dub {\it star}- and {\it diamond}-shaped graph states~\cite{footnote}. Both can be synthesised from the general graph state of $N+1$ qubits
\begin{equation}
\ket{G_{N+1}} = \prod_{j,k}\hat C(\phi_{j,k}) \left(\bigotimes^{N+1}_{l=1} \ket{+}_l\right),
\end{equation}
where we have introduced the controlled-phase gate $\hat C(\phi_{j,k})$ ($\phi_{j,k}$ is a real phase) acting on the qubit pair $(j,k)$ as 
\begin{equation}
\hat C(\phi_{j,k})=\ket{0}\bra{0}_{j}\otimes\openone_k+\ket{1}\bra{1}_j\otimes
\begin{pmatrix}
1&0\\
0&e^{i{\phi_{j,k}}}
\end{pmatrix}_k.
\end{equation}
The choice of $\phi_{j,k}=\pi$ for any pair of qubits results in a standard cluster state~\cite{Briegel}.

In order to generate the graph states of interest, we perform a controlled-phase gate with phase $\phi$ between the qubit of the register that is chosen as the system (which we assume to be qubit $1$ for simplicity) and its neighbouring qubits. On the other hand, we apply a controlled-phase gate with phase $\theta$ between the pairs of qubits of the environment. This results in 
\begin{equation}
\ket{G_{N+1}} =\prod_{j=2}^{N-2}\hat C(\theta_{j,j+1})\prod_{k=2}^{N}\hat C(\phi_{{\sys},k}) \left( \ket{+}_{\sys} \bigotimes^N_{l=2}\ket{+}_l\right)
\end{equation}
with $\phi_{{\sys},k}=\phi$ and $\theta_{j,j+1}=\theta$ for any choice of indices. By changing these values we change the strength of the correlations between the qubits.
%
In particular, for $\phi=\pi$ and $\theta=0$, which correspond to the star-shaped graph state, no interaction among the elements of the environment is set, and thus no intra-environment correlation can be set. This corresponds to the standard assumptions in the QD picture. On the other hand, a non-zero value of $\theta$ corresponds to a diamond-shaped graph state of $\theta$-dependent degree of correlations among the ${\cal E}_j$'s.

We analyze the mutual information shared between the system and the environment as a function of fractions of the environment for such systems. The theoretical predictions are shown as the lines in Fig. \ref{fig:graph01}c-d. 
There is a strongly different behaviour between the Star and the Diamond cluster state: the former has a substantially constant mutual information, which is sign of the presence of Darwinism in the system, the latter has a 
growing mutual information which denotes Darwinism's disruption. We note that for both the configurations considering the whole environment leads to complete retrival of the information $I_{\sys\frag}=2S(\rho_{\sys})$. Consequently the star-shaped graph state shows evidence of objective reality, while the diamond-shaped one does not.


\begin{figure*}[t!]
	\includegraphics[width=0.99\textwidth]{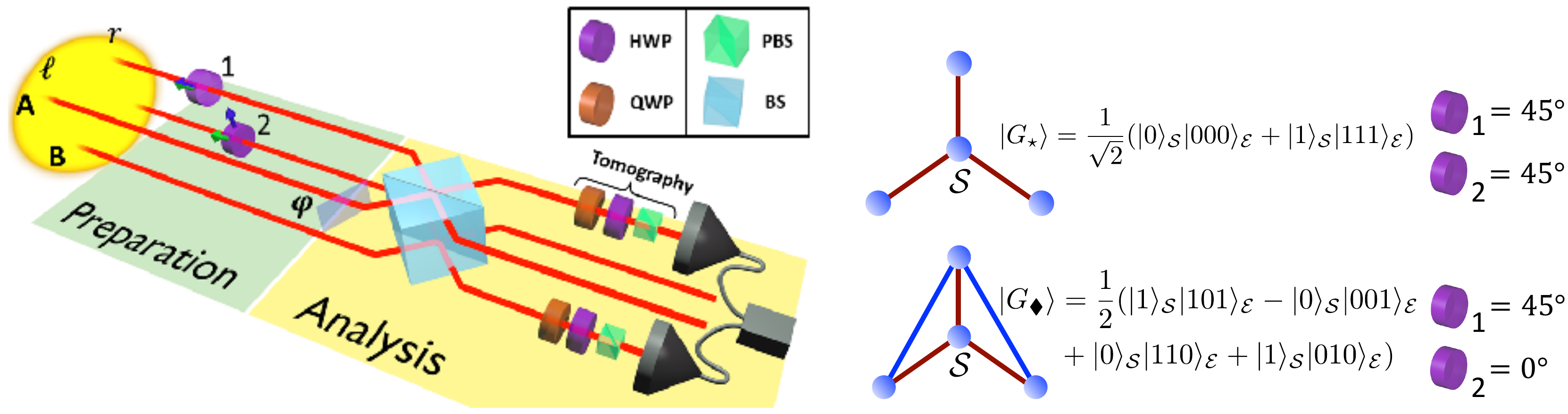}
	\caption{Experimental scheme for the synthesis of diamond- and star-shaped graph states. In the figure, HWP stands for a half waveplate, QWP is a quarter waveplate, PBS is a polarizing beamsplitter, BS is a beamsplitter. In the preparation stage, the resource states are built. The yellow halo represents the path-polarization two-photons four-qubit hyperentangled source described in Refs.~\cite{LSACiampini, Barbieri}. 
\emph{Star-shaped graph state}: the source generates the state $\ket{HH}_{AB}\otimes(\ket{\ell r}+\ket{r\ell})_{AB}$ ; HWP1 and HWP2 are placed at $45\si{\degree}$  on modes $r_A$ and $ r_B$ (green arrows).
\emph{Diamond-shaped graph state:} the source generates $(\ket{HH}+\ket{VV})_{AB}\otimes(\ket{\ell r}+\ket{r\ell})_{AB}$; the half waveplate HWP1 on mode $r_A$ is oriented at $45\si{\degree}$, while HWP2 on mode $r_B$ is at $0\si{\degree}$ (blue arrows). In the analysis stage, the path qubits are analyzed through a phase shifter ($\varphi$) and a BS, the polarization qubits are analyzed through a standard tomographic setup. Interferometric filters select degenerate photons centered at $\lambda=710$nm with a $6$nm bandwitdh, and coincidence counts between modes $r_A$ and $l_B$ are measured using single- photon counting modules (SPCMs) in a time window of $\approx 9$ns. Rates of $\approx 500$ coincidences/sec are experimentally observed.}
	\label{fig:exp_setup}
\end{figure*}

\emph{Experimental synthesis of star- and diamond-shaped graph states. --} We now illustrate the experimental procedure used to engineer representatives of the two classes of states discussed above. Our approach is based on the use of a well-consolidated and tested source of four-qubit path-polarization photonic hyperentangled cluster states~\cite{Barbieri,Cinelli}. 

A double-passage scheme through a non linear type-I $\beta$ Barium-Borate (BBO) crystal generates a two-photon state, entangled in polarization, while a four-hole mask, symmetrically placed over the centre of the photons emission cone, generates entanglement between the paths of the photon pairs. The state generated by this source can be written as
\begin{equation}
\ket{\Xi}=\frac{1}{\sqrt{2}}(\ket{HH}_{12}+\ket{VV}_{12})\otimes \frac{1}{\sqrt{2}}(\ket{\ell r}_{34}+\ket{r \ell}_{34}).
\label{eq:HE}
\end{equation}
Here $\ket{H}$ and $\ket{V}$ are horizontal and vertical polarization states of the two photons, respectively, while $\ell$ and $r$ are left and right path of the photons through the mask. The label $j=1,\dots,4$ identify the logical qubits at hand.

The full form of a four-qubit star-shaped graph state is
\begin{equation}
\ket{G_{\star}}=\left[\prod^4_{j=2}\hat C(\phi_{{\sys},j})\right] \ket{++++}_{1234},
\end{equation}
where $\phi_{{\sys},j}=\pi~\forall j=2,..,4$. It can be easily seen that $\left(\bigotimes^4_{j=2} \hat {\cal H}_j\right) \ket{G_{\star}}=\ket{GHZ_4}$ with $\ket{GHZ_4}=(\ket{0000}_{1234}+\ket{1111}_{1234})/\sqrt2$ a four-qubit Greenberger-Horne-Zeilinger (GHZ) state, and $\hat{\cal H}_j$ a Hadamard gate applied to qubit $j$. As local operation on single qubits do not change the mutual information within the state, we can equivalently analyze the GHZ state. This is retrieved from the  experimental state in Eq.~(\ref{eq:HE}) by selecting only one of the cones of the emission from the BBO crystal and introducing two half-wave plates (HWPs) at $45\degree$ on the paths of modes $r_A$ and $r_B$. The resulting state thus becomes
$\ket{G_{\star}}_{exp}=\frac{1}{\sqrt{2}}(\ket{HV\ell r}_{1234}+\ket{VHr\ell}_{1234})$.
The logical encoding $\ket{H/V}\rightarrow\ket{0/1}$, $\ket{\ell / r}\rightarrow\ket{0/1}$ gives us the state $(\ket{0101}+\ket{1010})/\sqrt2$, which is locally equivalent to the GHZ state above.

As for the four-qubit diamond-shaped graph state, we can proceed as follows. The state is written as
$\ket{G_\blacklozenge}=\hat C(\theta_{23})\hat C(\theta_{34})\ket{G_{\star}}$
with $\theta_{23}=\theta_{34}=\pi$. A straightforward manipulation shows that 
\begin{equation}
\begin{aligned}
& \text{Swap}_{2,3} \mathcal{H}_1 \otimes (\sigma_x {\cal H})_2 \otimes {\cal H}_3 \otimes (\sigma_z {\cal H})_4 \ket{G_\blacklozenge}\\ &=\frac{1}{2}(-\ket{0001}+\ket{0110}+\ket{1010}+\ket{1101})_{1234}
\label{eq:diam01}
\end{aligned}
\end{equation}
with $\text{Swap}_{2,3}$ the swap gate applied to qubits $2$ and $3$.  In order to build the state through our experimental setup, we start from  Eq.~(\ref{eq:HE}) and we 
introduce a HWP at $45\degree$ 
over mode $r_A$ and one at $0\degree$ 
over mode $r_B$. The result is 
$\ket{G_{\blacklozenge}}_{exp}=\frac{1}{2}[-(\ket{HH}-\ket{VV})_{12}\ket{\ell r}_{34}+(\ket{HV}+\ket{VH})_{12}\ket{r\ell}_{34}]$,
which is equivalent to Eq.~(\ref{eq:diam01}). 

\emph{Experimental assessment of QD. --} 
\begin{figure}[b!]
	\includegraphics[width=0.99\columnwidth]{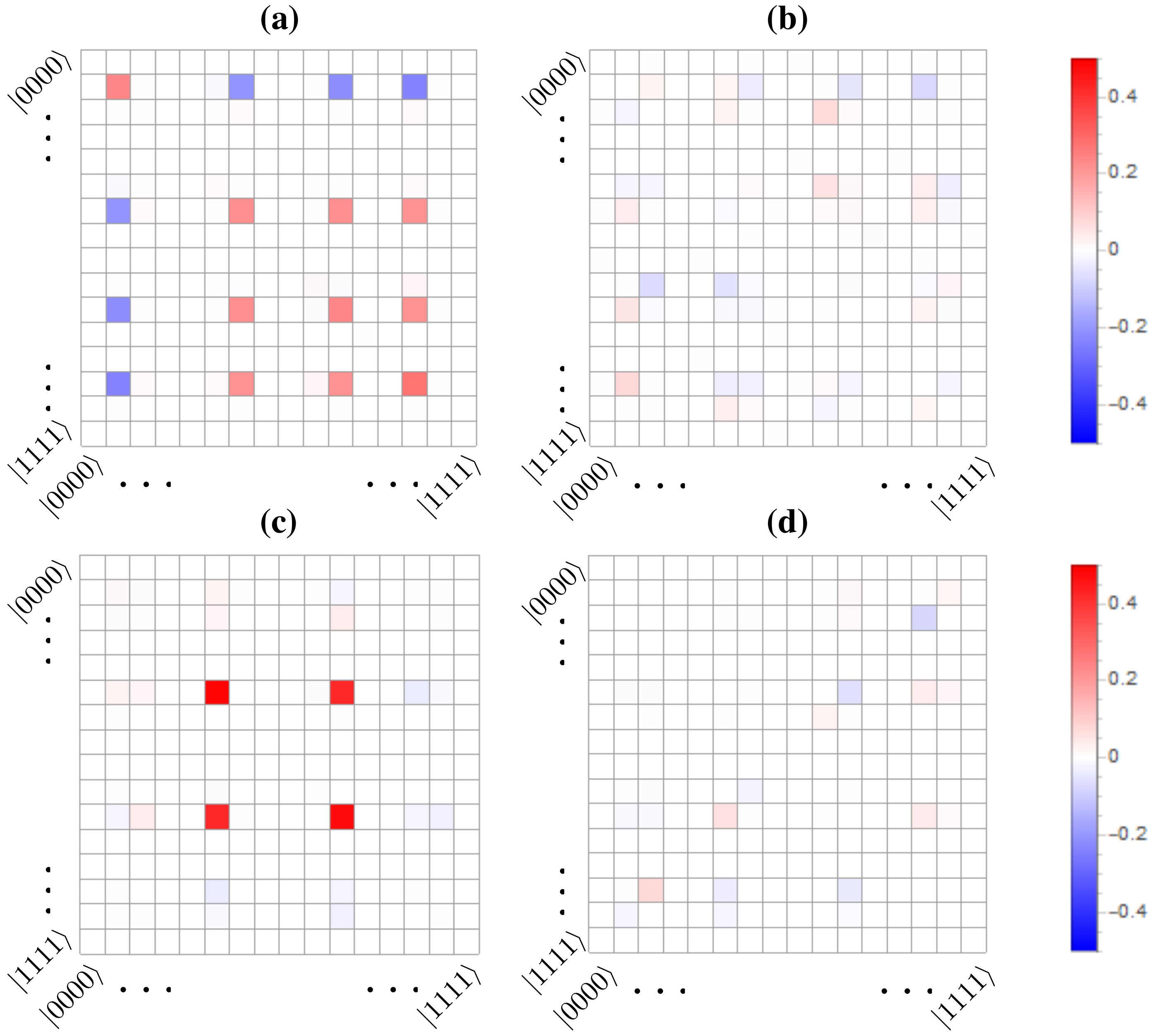}
	\caption{Experimental density matrix of the diamond- (top) and star-shaped (bottom) graph state. Panels {\bf (a)} and {\bf (c)} [Panels {\bf (b)} and {\bf (d)}] show the real [imaginary] parts of the density matrix entries.}
	\label{fig:tomo}
\end{figure}
As mentioned previously, the mutual information between $\sys$ and a growing-size environmental fragment ${\cal F}$ allows us to assess the phenomenology of redundant encoding due to the studied system-environment interaction. A possible approach to perform such a study is the {\it evaluation} of $I_{\sys\frag}$ over the reductions of the experimentally reconstructed $\sys$-$\env$ density matrix. As a hyperentanglement-based approach allows for the fully independent control over the four qubits of our resource state, it represents an optimal platform for this assessment. In Fig. \ref{fig:tomo} we present the full state tomographies that have been experimentally determined to evaluate the quality of the generated states. We obtained a fidelity of $F_\star = (91.0\pm0.7)\%$ for the star state, and $F_\blacklozenge =(91\pm1)\%$ for the diamond state which highlights the good overall quality of our states.  

\begin{figure}[b!]
\includegraphics[width=\columnwidth]{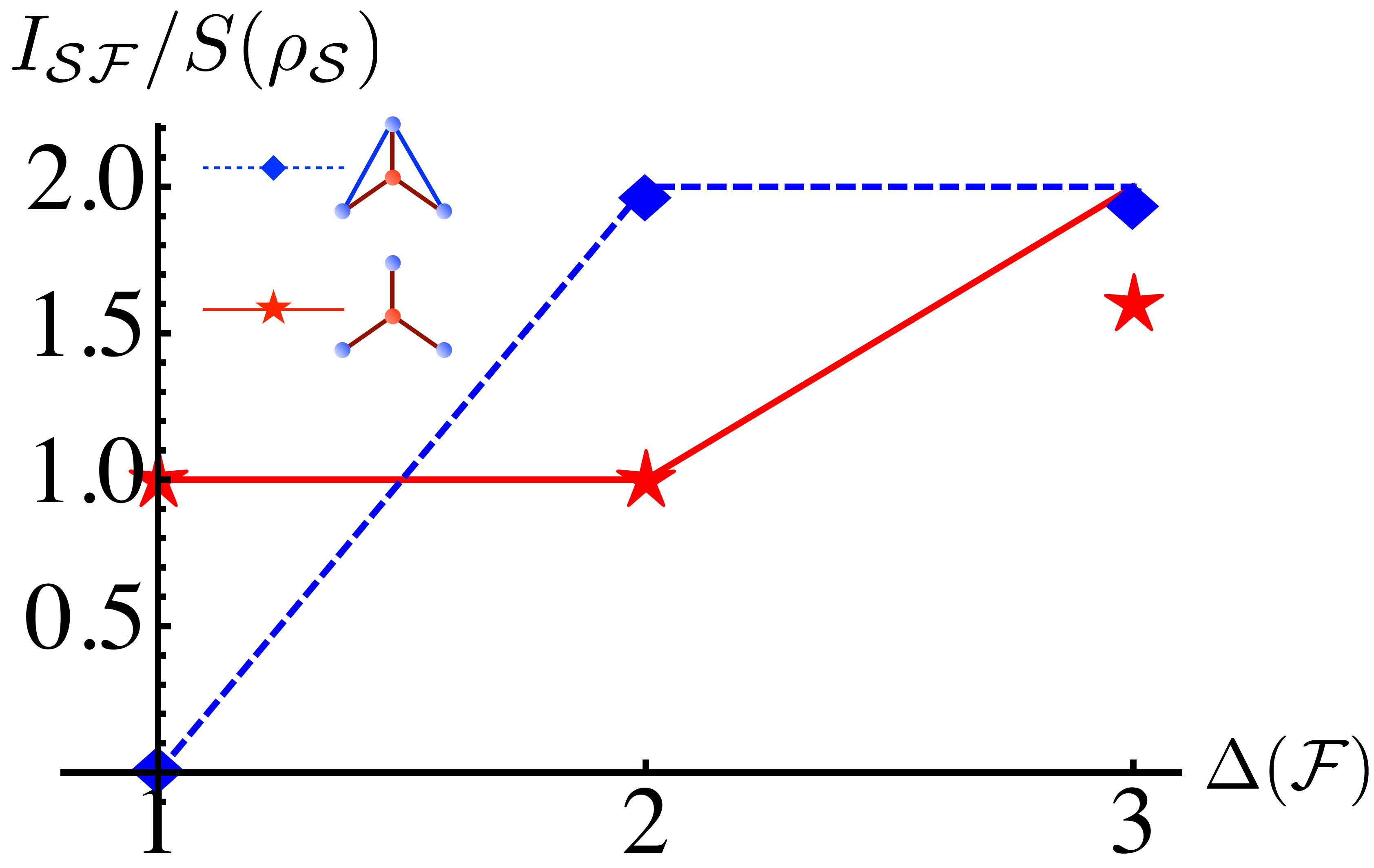}\\
	\caption{Experimental mutual information between system and fractions of the environment, evaluated with the decomposition of the density matrix, as described in the main text. Red stars are for a star-shaped graph state, while the blue diamonds are for the diamond-shaped one. The red-solid and blue-dashed lines represent theoretical predictions. 
The error bars, representing a one-sigma confidence on the experimental point, are smaller than the dimensions of the symbols and are evaluated by considering Poissonian statistics on the photonic coincidence counts.}
	\label{fig:exp_results}
\end{figure}


However, an experimentally more direct estimation of the mutual information is possible adapting the methodology put forward by some of us in Ref.~\cite{Chiuri}, which is based on the decomposition of the mutual information between $\sys$ and a given fragment $\frag$ in terms of multi-qubit correlators of the elements of the set of operators $\{\bm\sigma\}$ with $\sigma_0=\openone$ and $\sigma_{1,2,3}$ the $x$, $y$, and $z$ Pauli matrix, respectively. In details, let us define the four-point correlation functions evaluated over the density matrix $\rho^k_{exp}$ ($k=\star,\blacklozenge$) describing the state of our four-qubit resource
\begin{equation}
{\cal C}^k_{\alpha\beta\gamma\delta}=\Tr\left[\rho^k(\sigma_\alpha\sigma_\beta\sigma_\gamma\sigma_\delta)\right]~~~~~~(\alpha,\beta,\gamma,\delta=0,\dots,3),
\end{equation}
where we have omitted the symbol of tensor product for easiness of notation. Such correlators can be easily measured in our experimental setup, similarly to the procedure used to reconstruct the density matrix, by means of path projections, obtained by introducing a beam splitter (BS) over the four output modes of the mask \cite{Barbieri}, and polarization projections, obtained by a standard tomographic setup made by a quarter wave-plate (QWP), a HWP and a polarizing BS (PBS). Coincidence counts are measured by avalanche photodiodes in a gate of $\sim 9ns$. Suitable combinations of such correlation functions allow us to reconstruct the element $(\rho^k)_{ij}$ of the density matrix as
\begin{equation}
(\rho^k)_{ij}=\sum^3_{\alpha,\beta,\gamma,\delta=0} a^{ij}_{\alpha\beta\gamma\delta} {\cal C}^k_{\alpha\beta\gamma\delta}~~~~~~(i,j=1,16)
\end{equation}
with $\{a^{ij}_{\alpha\beta\gamma\delta}\}$ as set of (in general complex) numbers. The $\sys$-$\frag$ mutual information can then be cast as a function of such correlation functions. 
For instance, for the case of a star-shaped graph state we have $\rho^\star=\ket{G_{\star}}\bra{G_\star}=P\ket{0101}\bra{0101}+(1-P)\ket{1010}\bra{1010}+(C\ket{0101}\bra{1010}+h.c.)$ with
\begin{equation}
\begin{aligned}
P&=\left[{\cal C}_{0000}+{\cal C}_{3333}-{\cal P}({\cal C}_{0003})-{\cal P}({\cal C}_{0033})+{\cal P}({\cal C}_{0333})\right]/16,\\
C&=\left[{\cal C}_{1111}+{\cal C}_{2222}-i{\cal P}({\cal C}_{1112})-i{\cal P}({\cal C}_{1222})+{\cal P}({\cal C}_{1122})\right]/16,
\end{aligned}
\end{equation}
where ${\cal P}({\cal C}_{\alpha\beta\gamma\delta})$ is the operator that performs the sum over the correlators obtained by permutation of the  
indices $\alpha$, $\beta$, $\gamma$, $\delta$. The mutual information $I^{\Delta(\frag)}_{\sys\frag}$ for $\Delta({\frag})=1,2,3$ is then given by the following functions of $P$ and $C$
\begin{equation}
\begin{aligned}
I^1_{\sys\frag}&=I^2_{\sys\frag}=-P_r\log(P_r)-(1-P)_r\log(1-P)_r,\\
I^3_{\sys\frag}&=\sum_{k=\pm} f^k_r\log f^k_r-2 [P_r\log P_r+ (1-P)_r\log(1-P)_r]
\end{aligned}
\end{equation}
with $f^\pm=(2P-1\pm\sqrt{4|C|^2+(1-2P)^2})/2$ and the subscript $r$ which stands for the real part of the corresponding function. Therefore, the experimental estimate of the mutual information can be performed by determining both $P$ and $C$ over the experimental state $\rho^\star_{exp}$. 
In the case of $\rho_{exp}^\star$ this method only requires $32$ measurements, which compares very favourably with respect to the $6^4=1296$ measurements that are required for a full quantum state tomography. A similar analysis can be performed for the state $\rho^{\blacklozenge}_{exp}$. The experimental estimate of the mutual information for both the resource states are reported in Fig.~\ref{fig:exp_results}, which demonstrates the striking differences in the behavior associated with the two state configurations~\cite{comment}. The presence of correlations between elements of the environment, as it is the case for a diamond-shaped graph state, results in an enhanced ability of the observers to gather information on the state of the system, which becomes maximum already when only $75\%$ of the environmental elements is queried. Despite the small size of the resource that has been synthesised in our experiment, this feature emerges strikingly, {\it de facto} validating the expected trend of the mutual information displayed in Fig.~\ref{fig:graph01}. A comment is in order: contrary to expectation and examples given in this respect~\cite{Galve}, the observed breakdown of QD does not result from the emergence of non-Markovian dynamics due to the enforced interactions among the elements of the (finite-size) environment. In fact, the non-Markovian character of the system's dynamics in the diamond-shaped graph state is less pronounced than the one resulting from the star-shaped configuration. This suggests a more convoluted than anticipated relationship between memory-bearing environmental mechanisms and the emergence of objective reality. The establishment of such a link goes beyond the scopes of this work and will be the focus of further investigations.

\emph{Discussion.-- } We have assessed the effect that strong intra-environment correlations have over the emergence of QD in a controlled simulator of quantum open-system dynamics. We have shown that a simple graph configuration is able to encompass fundamental alterations, with respect to the behaviour expected to arise from the typical QD paradigm~\cite{Zurek2,Zurek3,Zurek4}, in the way information on the state of a quantum system is shared by the elements of an environment. Remarkably, the onset of such modifications to the QD phenomenology occurs already at a very small size (i.e. they are not emergent), which has enabled their experimental verification in a four-qubit photonic cluster state. 

While representing a rare instance of experimental case-study on the emergence of objective reality (or the lack thereof), our work also highlights the complexity of this phenomenon, and its fragility with respect to critical dependence of the actual system-environment dynamics from the prescriptions of QD.

\emph{Acknowledgements.--} This work was partially supported by the project FP7-ICT-2011-9-600838 (QWAD Quantum Waveguides Application and Development; www.qwad-project.eu). MP acknowledge the DfE-SFI Investigator Programme (grant 15/IA/2864), the H2020 Collaborative Project TEQ (grant 766900), and the Royal Society for financial support.


\begin{thebibliography}{99}
\bibitem{footnote} A completely symmetric state resulting from the addition of a ring-shaped graph to the star-shaped one leads to the vanishing of quantum correlations in the environment. Therefore, such structure is equivalent to the star-shaped graph state, as far as QD is concerned.
\bibitem{comment} The differences between theoretical predictions and experimental estimates of the mutual information can be explained by considering the effect of unwanted non-zero elements in the reconstructed experimental density matrixes.
\end{thebibliography}
\end{document}